\documentclass[11pt]{article}
\usepackage{moriond}
\usepackage{amsmath}
\usepackage{axodraw}

\def\mco{\multicolumn}

\def\vpj{\mbox{${\varphi^\dag i\,\raisebox{2mm}{\boldmath ${}^\leftrightarrow$}\hspace{-4mm} D_\mu\,\varphi}$}}
\def\vpjt{\mbox{${\varphi^\dag i\,\raisebox{2mm}{\boldmath ${}^\leftrightarrow$}\hspace{-4mm} D_\mu^{\,I}\,\varphi}$}}

\def\be{\begin{equation}}
\def\ee{\end{equation}}
\def\bea{\begin{eqnarray}}
\def\eea{\end{eqnarray}}

\newcommand{\Photo}{}

\begin{document}
\vspace*{4cm} 

\title{ $l^J \to l^I \gamma$ in the Standard Model with general
  dimension 6 terms }

\author{Saereh Najjari \footnote{saereh.najjari@fuw.edu.pl}}

\address{Institute of Theoretical Physics, University of Warsaw,
  00-681 Warsaw, Poland}

\maketitle\abstracts{We investigate the possibility of lepton flavor
  violation in the decays $\mu\to e\gamma$, $\tau \to \mu\gamma$ and
  $\tau \to e\gamma$ as well as electric dipole moment and anomalous
  magnetic moment of charged leptons in the extension of the Standard
  Model with the most general set of gauge-invariant operators up to
  dimension~6. }

\section{Introduction}
The Standard Model (SM) of strong and electroweak interaction is a
theory that successfully explains and predict the elementary particle
phenomenology.  It is considered to be a low energy approximation of a
more fundamental theory.  In the SM, the lepton number of each family
(lepton flavor) is conserved.  On the other hand, the discovery of
flavor violation would provide a possible hint of New Physics beyond
the SM. So far, neutrino oscillations are the only example of lepton
flavor violation (LFV) that has been observed.

It is possible to extend the SM by keeping its gauge symmetry, the
particle content and the pattern of spontaneously symmetry breaking by
adding new effective operators:
\bea
{\cal L}_{\rm SM} & = & {\cal L}_{\rm SM}^{\rm(4)}+ \frac{1}{\Lambda}
\sum_{k}{{C_k^{(5)}}Q_k^{(5)} }+\frac{1}{{\Lambda}^2}
\sum_{k}{{C_k^{(6)}}Q_k^{(6)} }+\mathcal{O}(\frac{1}{{\Lambda}^3}
) \label{SM}
\eea
where $\Lambda$ is large, of the order of the scale of New Physics.
$Q_k^{(n)}$ denote operators of dimension $n$ and $C_k^{(n)}$ stand
for corresponding dimensionless coupling constants (Wilson
coefficients). Once the underlying high-energy theory is specified,
all the coefficient $C_k^{(n)}$ can be determined by integrating out
the heavy fields.
\begin{table}[t]
\caption{Lepton Flavor Violating dimension-6 operators. }
\label{tab:exp}
\vspace{1cm}
\begin{center}
\begin{tabular}{|c|c|c|c|c|c|}
\hline\hline
\mco{2}{|c|}{$\ell\ell\ell\ell$}   &\mco{2}{|c|}{$\ell\ell X\varphi$}&
\mco{2}{|c|}{$\ell\ell\varphi^2 D ~and~ \ell \ell \varphi^3$}

\\ \hline &&&&&\\
$Q_{\ell\ell}$  & $({\bar{\ell_i}} \gamma_\mu \ell_j)(\bar \ell_k \gamma^\mu \ell_l)$ & $Q_{eW}$               & $(\bar \ell_i \sigma^{\mu\nu} e_j) \tau^I \varphi W_{\mu\nu}^I$ &
$Q_{\varphi \ell}^{(1)}$      & $(\vpj)({\bar{ \ell_i}} \gamma^\mu \ell_j)$\\$Q_{ee}$& $(\bar e_i \gamma_\mu e_j)(\bar e_k \gamma^\mu e_l)$&$Q_{eB}$ & $(\bar \ell_i \sigma^{\mu\nu} e_j) \varphi B_{\mu\nu}$ &$Q_{\varphi \ell}^{(3)}$ &$(\vpjt)(\bar \ell_i \tau^I \gamma^\mu \ell_j)$\\
$Q_{\ell e}$ & $(\bar \ell_i \gamma_\mu \ell_j)(\bar e_k \gamma^\mu e_l)$  &
       &  &
$Q_{\varphi e}$            & $(\vpj)(\bar e_i \gamma^\mu e_j)$\\&& && $Q_e^{\varphi^3}$ & $(\varphi^\dag \varphi)(\bar \ell_i e_j \varphi)$
\\
\hline \hline
\mco{6}{|c|}{$\ell\ell qq$} \\ \hline
$Q_{\ell q}^{(1)}$&$(\bar \ell_i \gamma_\mu \ell_j)(\bar q_k \gamma^\mu q_l)$ &$Q_{\ell d}$&$(\bar \ell_i \gamma_\mu \ell_j)(\bar d_k \gamma^\mu d_l)$&$Q_{\ell u}$ &$(\bar \ell_i \gamma_\mu  \ell_j)(\bar u_k \gamma^\mu  u_l)$\\
$Q_{\ell q}^{(3)}$ & $(\bar \ell_i \gamma_\mu \tau^I \ell_j)(\bar q_k \gamma^\mu \tau^I q_l)$&
$Q_{ed}$ & $(\bar e_i \gamma_\mu e_j)(\bar d_k\gamma^\mu d_l)$&
$Q_{eu}$  & $(\bar e_i \gamma_\mu e_j)(\bar u_k \gamma^\mu u_l)$\\
$Q_{eq}$  & $(\bar e_i \gamma^\mu e_j)(\bar q_k \gamma_\mu q_l)$ &
$Q_{\ell edq}$ & $(\bar \ell_i^a e_j)(\bar d_k q_l^a)$ &
$Q_{\ell equ}^{(1)}$ & $(\bar \ell_i^a e_j) \varepsilon_{ab} (\bar q_k^b u_l)$ \\
&& &&
$Q_{\ell equ}^{(3)}$ & $(\bar \ell_i^a \sigma_{\mu\nu} e_a) \varepsilon_{ab} (\bar q_k^b \sigma^{\mu\nu} u_l)$ \\
\hline\hline

\end{tabular}
\end{center}
\end{table}

In Table~\ref{tab:exp}, we collected the independent dimension-6
operators which could contribute to LFV process in the charged lepton
sector at tree level or at one-loop level (there is a unique operator
of dimension 5 which gives Majorana mass to neutrinos, but its
contribution to LFV in charged lepton sector is negligible). The
complete set of dimension 5 and 6 operators can be found in
\cite{Grzadkowski:2010es}.

\section{Radiative Lepton Decays}

\begin{figure}[htb]
\begin{center}
\begin{tabular}{cccc}
\begin{picture}(90,70)(10,0)
\Photon(50,0)(50,50){3}{4} \Text(45,55)[c]{$\gamma^{\mu}$}
\ArrowLine(60,25)(60,40) \Text(67,35)[l]{$q=p_J-p_I$}
\ArrowLine(20,0)(40,0)
\Text(10,0)[l]{$\ell^J$}
\Text(30,10)[c]{$p_J$}
\ArrowLine(60,0)(80,0)
\Text(90,0)[c]{$\ell^I$}
\Text(75,10)[c]{$p_I$}
\GCirc(50,0){10}{0.5}
\end{picture}
&
\begin{picture}(120,70)(10,0)
\Photon(90,0)(90,50){3}{4}
\Text(85,55)[c]{$\gamma^{\mu}$}
\ArrowLine(20,0)(40,0)
\Text(10,0)[l]{$\ell^J$}
\ArrowLine(90,0)(110,0)
\Text(120,0)[c]{$\ell^I$}
\GCirc(50,0){10}{0.5}
\Vertex(90,0){2}
\ArrowLine(60,0)(90,0)
\Text(70,10)[l]{$\ell^I$}
\end{picture}
&
\begin{picture}(130,70)(20,0)
\Photon(50,0)(50,50){3}{4}
\Text(45,55)[c]{$\gamma^{\mu}$}
\ArrowLine(30,0)(50,0)
\Text(20,0)[l]{$\ell^J$}
\ArrowLine(100,0)(120,0)
\Text(130,0)[c]{$\ell^I$}
\GCirc(90,0){10}{0.5}
\Vertex(50,0){2}
\ArrowLine(50,0)(80,0)
\Text(60,10)[l]{$\ell^J$}
\end{picture}
&
\begin{picture}(90,70)(10,0)
\Photon(50,0)(50,50){3}{4} 
\Text(45,55)[c]{$\gamma^{\mu}$}
\Text(55,12)[l]{$Z^0,\gamma^0,G^0$}
\ArrowLine(20,0)(50,0)
\Text(10,0)[l]{$\ell^J$}
\ArrowLine(50,0)(80,0)
\Text(90,0)[c]{$\ell^I$}
\GCirc(50,30){10}{0.5}
\Vertex(50,0){2}
\end{picture}
\\
\end{tabular}
\end{center}
\caption{Topologies of diagrams contributing to radiative decay
  $\ell^J\to \ell^I \gamma$.  \label{fig:llgen}}.
\end{figure}
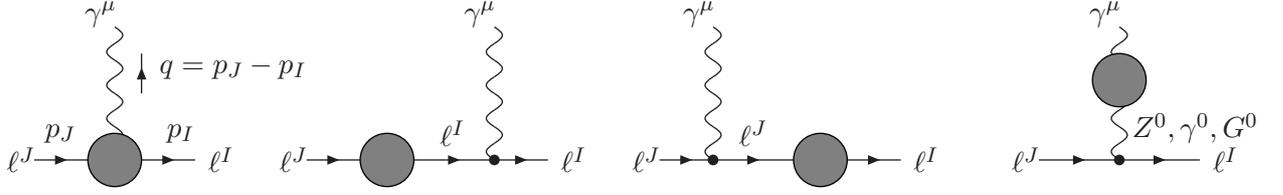

In this section we calculate the radiative lepton decays $\ell^J \to
\ell^I\gamma$ , where $I, J$ are flavor indices running from 1 to 3.
The general form of lepton-photon vertex can be written as:
\bea
V_{\ell \ell\gamma}^{IJ\;\mu} =
\frac{i}{\Lambda^2}[\gamma^\mu(F_{VL}^{IJ} P_L + F_{VR}^{IJ} P_R)
  + (F_{SL}^{IJ} P_L + F_{SR}^{IJ} P_R) q^\mu + (F_{TL}^{IJ}
  \sigma^{\mu\nu} P_L + F_{TR}^{IJ} \sigma^{\mu\nu} P_R) q_\nu]
\eea
Only the form-factors $F_{TL}$ and $F_{TR}$ contribute to $\ell^J \to
\ell^I \gamma$ decay. The operators $Q_{eW}$ and $Q_{eB}$ can give
contribution to these decays at tree-level, it can be written as:
\bea
F_{TR}^{IJ} = F_{TL}^{IJ\star} = v\sqrt{2} \left(c_W C_{eB}^{IJ} - s_W
C_{eW}^{IJ} \right)
\label{eq:phtree}
\eea
If these Wilson coefficient, $C_{eB}$ and $C_{eW}$, vanish or are very
small (in every renormalizable theory they can be only
loop-generated), other operators can contribute to the lepton to
lepton and photon decay at one loop level (see Fig.~\ref{fig:llgen}).
The list of Feynman rules arriving from those operators and the one
loop diagrams contributing to these decays are given
in~\cite{Crivellin:2013hpa}.  The branching ratio for the $\ell^J \to
\ell^I\gamma$ (with $J>I$) is given by:
\bea
{\cal B} \left[\ell^{J} \to \ell^{I} \gamma \right] \,=\,
\dfrac{m_{\ell_J}^3}{16\pi\Lambda^4 \, \Gamma_{\ell_J}} \left(
\left|F^{IJ}_{TR} \right|^{2}+ \left|F^{IJ}_{TL} \right|^{2} \right )
\,,
\label{Brmuegamma}
\eea
Where $\Gamma_{\ell_J}$ is the total decay width of decaying lepton.

The finite 1-loop results for $F_{TL}$, $F_{TR}$ form-factors are:
\bea
F_{TL}^{IJ} &=& {2e \over (4\pi)^2} \left[{2 [ ( C_{\phi l}^{(1)IJ} +
      C_{\phi l}^{(3)IJ} ) m_I (1+s_W^2) - C_{\phi e}^{IJ} m_J
      (\frac{3}{2} - s_W^2)] - 5 m_I C_{\phi l}^{(3)IJ} \over 3} +
  \sum_{K=1}^3 C_{\ell e}^{IKKJ} m_K\right] \nonumber\\
F_{TR}^{IJ} &=& {2e \over (4\pi)^2} \left[ {2 [ ( C_{\phi l}^{(1)IJ} +
      C_{\phi l}^{(3)IJ} ) m_J (1+s_W^2) - C_{\phi e}^{IJ} m_I
      (\frac{3}{2} - s_W^2)] - 5 m_J C_{\phi l}^{(3)IJ} \over 3} +
  \sum_{K=1}^3 C_{\ell e}^{KJIK} m_K\right] \nonumber\\
\label{eq:ft}
\eea

Operator $Q_{\ell equ}^{(3)}$ generates divergent contribution, which
can be understood and renormalized in a complete theory of New Physics
(see discussion in ref.~\cite{Crivellin:2013hpa}). As such term
contains quark fields and can be constrained from hadronic decays, we
do not consider it in our numerical analysis.

For $I=J$, results presented in Eq.~(\ref{eq:ft}) can be directly used
to calculate the leptonic electric dipole moments (EDM),
\bea
d_{l_J}= {-1 \over \Lambda^2} {\rm Im}[F_{TR}^{JJ}]
\eea
as well as their anomalous magnetic moments:
\bea
a_{\ell_ J}={2m_{\ell_J}\over{e\Lambda^2}}{\rm Re}[F_{TR}^{JJ}]\;.
\eea

\section{Numerical Analysis}
\label{sec:numal}

\begin{table}[t]
\caption{Experimental upper limits on the branching ratios of the
  radiative lepton decays.}
   \label{tab:expBr}
   \centering \vspace{0.8cm}
\renewcommand{\arraystretch}{1.2}
  \begin{tabular}{@{}|c|c|}
 \hline
Process & Experimental bounds \\ \hline \hline
${\cal B}\left[ \tau \to \mu \gamma \right] $ & ~$\leq \, 4.4\times
10^{-8}$ ~\cite{Aubert:2009ag,Hayasaka:2007vc} \\
\hline
${\cal B}\left[ \tau \to e \gamma \right] $ & ~$\leq \, 3.3 \times
10^{-8}$~ \cite{Aubert:2009ag} \\
\hline
$ {\cal B}\left[ \mu \to e \gamma \right] $ & ~$\leq \, 5.7 \times
10^{-13}$~\cite{Adam:2013mnn}
\\
\hline \hline
\end{tabular}
 \label{table:RLFVdecays}
\end{table}

The analytical results listed in previous Section can be compared with
experimental bounds on radiative lepton charged decays given in
Table~\ref{tab:expBr}.  As they are dominated by tree-level
contributions from $Q_{eB},Q_{eW}$, denoting $C_\gamma^{IJ}\equiv c_W
C_{eB}^{IJ}-s_W C_{eW}^{IJ}$ one gets
\bea
\sqrt{| {C_\gamma^{12}}|^2+| C_\gamma^{21}|^2}&\leq&
2.45\times10^{-10}({{\Lambda}\over 1~TeV})^2{\sqrt{Br[\mu \to
      e\gamma]\over{5.7\times10^{-13}}}}\;,\nonumber\\
\sqrt{| {C_\gamma^{13}}|^2+| C_\gamma^{31}|^2} &\leq
&2.35\times10^{-6}({{\Lambda}\over {1~TeV}})^2{\sqrt{Br[\tau \to
      e\gamma]\over{3.3\times10^{-8}}}}\;,\nonumber\\
\sqrt{| {C_\gamma^{23}}|^2+| C_\gamma^{32}|^2}&\leq&
2.71\times10^{-6}({{\Lambda}\over {1~TeV}})^2{\sqrt{Br[\tau \to
      \mu\gamma]\over{4.4\times10^{-8}}}}\;.
\eea
By neglecting the small lepton mass ratios and taking into account
that the Wilson coefficient of the 4-lepton and $Z^0$-lepton vertices
are real in flavor conserving case, we find following expressions for
the EDMs (the current experimental bound are given in
Table~\ref{tab:expEDM}):
\bea 
d_e&=& -6.86\times 10^{-18}
{\rm Im}[2\times10^{-5}C_{le}^{3113}+C_\gamma^{11}]({{1~TeV}\over{\Lambda}})^2
~e ~cm\;,\nonumber\\
d_\mu&=& -6.86\times 10^{-18}
{\rm Im}[2\times10^{-5}C_{le}^{3223}+C_\gamma^{22}]({{1~TeV}\over{\Lambda}})^2~e~
cm\;,\nonumber\\
d_\tau&=& -6.86\times 10^{-18}
{\rm Im}[C_\gamma^{33}]({{1~TeV}\over{\Lambda}})^2 ~e~ cm
\eea
and for anomalous magnetic moments:
\bea
a_e&=&1.17\times 10^{-6}{\rm
  Re}[2\times10^{-5}C_{le}^{3113}+C_\gamma^{11}]({{1~TeV}\over{\Lambda}})^2\;
,\nonumber\\
a_\mu&=&2.43\times 10^{-4}{\rm
  Re}[2\times10^{-5}C_{le}^{3223}+C_\gamma^{22}]({{1~TeV}\over{\Lambda}})^2\;
,\nonumber\\
a_\tau&=&4.1\times 10^{-3}{\rm Re}[10^{-5}\times({1.9C_{\Phi
      \ell}^{(1)33}}+2.0 C_{\ell e}^{3333}-1.4 C_{\Phi
    \ell}^{(3)33}-1.7C_{\Phi
    e}^{33})+C_\gamma^{33}]({{1~TeV}\over{\Lambda}})^2\;.
\eea 
The expression for $a_e$ can be compared with the difference between
experiment and theory for anomalous magnetic moment of
muon~\cite{Beringer:1900zz}:
\bea
\Delta a_\mu\equiv a_\mu^{exp}-a_\mu^{SM}=(287\pm 80)\times
10^{-11}\;.
\eea

\begin{table}[t]
\caption{Experimental upper bounds on electric dipole moments of the
  charged leptons.}
  \label{tab:expEDM}
\centering \vspace{0.8cm}
\renewcommand{\arraystretch}{1.2}
\begin{tabular}{|c| c| c| c| c|}
\hline \hline
EDMs & ${\left|d_{e}\right|}$ & ${|d_{\mu}|}$ & ${d_{\tau}}$ \\
\hline \hline
Bounds (${\rm e \, cm}$) & $8.7\times 10^{-29}$ \cite{Baron:2013eja}
& $1.9 \times 10^{-19} $ \cite{Bennett:2008dy} & $\in \left[ -2.5,
  \,0.8 \right] \times 10^{-17}$ \cite{Inami:2002ah} \\
\hline \hline
\end{tabular}

\label{tab:EDMs}
\end{table}

\section{Conclusions}

In this paper we discussed the tree-level and one-loop predictions for
radiative lepton decays, leptonic electric dipole moments and
anomalous magnetic moments of charged leptons in the SM extended with
all lepton flavor violating operators of dimension 6. We presented
compact analytical formulae for those processes in terms of Wilson
coefficients of new operators.  Numerical expressions based on such
formulae and on relevant experimental measurements, collected in
Section~\ref{sec:numal} can be used to obtain bounds on the magnitude
of Wilson coefficients depending on the scale of New Physics.

\section*{Acknowledgments}
This work has been done in collaboration with A. Crivellin and
J. Rosiek.  This research project has been supported by a Marie Curie
Initial Training Network of the European Community's Seventh
Framework Programme under contract number (PITN-GA-2009-237920-
UNILHC).


\section*{References}

\begin{thebibliography}{99}
\bibitem{Grzadkowski:2010es} B. ~Grz{\c a}dkowski, M. ~Iskrzy{\'n}ski,
  M. ~Misiak and J. ~Rosiek,
  JHEP {\bf 1010}, 085 (2010) [arXiv:1008. 4884 [hep-ph]].
\bibitem{Crivellin:2013hpa}
  A.~Crivellin, S.~Najjari and J.~Rosiek,
  arXiv:1312.0634 [hep-ph].
\bibitem{Aubert:2009ag}
  B.~Aubert {\it et al.}  [BaBar Collaboration],
  Phys.\ Rev.\ Lett.\  {\bf 104}, 021802 (2010)
  [arXiv:0908.2381 [hep-ex]].
\bibitem{Hayasaka:2007vc}
  K.~Hayasaka {\it et al.}  [Belle Collaboration],
  Phys.\ Lett.\ B {\bf 666}, 16 (2008)
  [arXiv:0705.0650 [hep-ex]].
\bibitem{Adam:2013mnn}
  J.~Adam {\it et al.}  [MEG Collaboration],
  Phys.\ Rev.\ Lett.\  {\bf 110}, 201801 (2013)
  [arXiv:1303.0754 [hep-ex]].
\bibitem{Beringer:1900zz}
  J.~Beringer {\it et al.}  [Particle Data Group Collaboration],
  Phys.\ Rev.\ D {\bf 86}, 010001 (2012).
\bibitem{Baron:2013eja}
  J.~Baron {\it et al.}  [ACME Collaboration],
  arXiv:1310.7534 [physics.atom-ph].
\bibitem{Bennett:2008dy}
  G.~W.~Bennett {\it et al.}  [Muon (g-2) Collaboration],
  Phys.\ Rev.\ D {\bf 80}, 052008 (2009)
  [arXiv:0811.1207 [hep-ex]].
\bibitem{Inami:2002ah}
  K.~Inami {\it et al.}  [Belle Collaboration],
  Phys.\ Lett.\ B {\bf 551}, 16 (2003)
  [hep-ex/0210066].

\end{thebibliography}
\end{document}